\documentstyle[12pt]{article}
\begin{document}

\begin{titlepage}

\begin{center}
{\bf Budker Institute of Nuclear Physics}
\end{center}

\vspace{1cm}

\begin{flushright}
{\bf BINP 95-69}
\end{flushright}

\vspace{1cm}

\begin{center}
{\bf NUCLEAR STRUCTURE CORRECTIONS\\ TO DEUTERIUM HYPERFINE
STRUCTURE AND LAMB SHIFT}
\end{center}

\begin{center}
I.B. Khriplovich\footnote{e-mail address: khriplovich@inp.nsk.su},
A.I. Milstein\footnote{e-mail address: milstein@inp.nsk.su},
S.S. Petrosyan\footnote{Novosibirsk University}
\end{center}
\begin{center}
Budker Institute of Nuclear Physics, 630090 Novosibirsk,
Russia
\end{center}

\bigskip

\begin{abstract}
The low-energy theorem for the forward Compton scattering is
generalized to the case of an arbitrary target spin. The
generalization is used to calculate the corresponding contribution to
the deuterium hyperfine structure.  The nuclear-structure corrections
are quite essential in this case due to the deuteron large size.
Corrections of this type calculated here remove the discrepancy
between the theoretical and experimental values of the deuterium
hyperfine splitting.  Explicit analytical result is obtained also for
the deuteron polarizability contribution to the Lamb shift.
\end{abstract}

\vspace{7cm}

\end{titlepage}

\section{Introduction}
The hyperfine (hf) splitting in deuterium ground state has been
measured with high accuracy. The most precise experimental result for
it was obtained with an atomic deuterium maser and constitutes
\cite{wr}
\begin{equation}
\nu_{exp}\,=\,327\,384.352\,522\,2(17)\, \mbox{kHz}.
\end{equation}
Meanwhile the theoretical calculation including higher order pure QED
corrections gives
\begin{equation}
\nu_{QED}\,=\,327\,339.27(7)\, \mbox{kHz}.\;\;\;\;\;\;\;\;\;\;\;\;\;\;\;
\end{equation}
The last number was obtained by using the theoretical result for the
hydrogen hf splitting from Ref. \cite{by}
$$1\,420\,451.95(14)\,\mbox{kHz}$$
which does not include proton structure and recoil radiative
correction, and combining it with the theoretical ratio of the hf
constants in hydrogen and deuterium from Ref. \cite{ra}
$$4.339\,387\,6(8)$$
based on the ratio of the nuclear magnetic moments and including the
reduced mass effect in $|\psi(0)|^2$.

In the present article the discrepancy
\begin{equation}\label{dif}
\nu_{exp}\,-\,\nu_{QED}\,=\,45\, \mbox{kHz}
\end{equation}
is removed by taking into account the effects due to the finite size
of deuteron. Such effects are obviously much larger in deuterium than
in hydrogen. One nuclear-structure contribution to the deuterium hf
splitting was pointed out long ago \cite{bo} by some intuitive
arguments, and then discussed in more detail in Refs.
\cite{low,los,gf}. Here we treat the deuteron finite-size effects in
a systematic way. Not only the old result is reproduced, but new
corrections are obtained, among them that generated by the deuteron
electric and magnetic form-factors.

To calculate some contributions to the deuterium hf structure we
generalize the low-energy theorem for the Compton scattering to an
arbitrary target spin.

One more subject considered in this paper is the contribution of the
deuteron polarizability to the deuterium Lamb shift. The fact that
this contribution is close to the accuracy attained in experiment was
pointed out in Refs. \cite{plh,pwh} where the effect was calculated
in the square-well approximation for the nuclear potential. This
correction was calculated then with separable nuclear potentials in
Ref. \cite{lr}. Here we obtain in the zero-range approximation a
closed analytical result for the deuteron polarizability contribution
to the Lamb shift.

\section{Low-energy theorem for the forward Compton scattering
on an arbitrary target}

According to the well-known low-energy theorem for the Compton
scattering on a spin 1/2 hadron \cite{lo,gg}, this amplitude is described by
the pole Feynman diagrams. We are interested here not in the
spin-independent Thomson amplitude which is of zeroth order in the
photon frequency $\omega$, but in the next, spin-dependent term of
the $\omega$ expansion. This contribution as well can be easily
obtained directly in the nonrelativistic approximation \cite{mi}. In
this approximation the electromagnetic vertex can be immediately
written for an arbitrary spin $\vec{s}$:
\begin{equation}\label{v}
\frac{e}{2m_p}\,\left\{\frac{Z}{A}\,(2\vec{p}\,+\,\vec{k})\,
+\,\frac{\mu}{s}\,i[\vec{s}\times\vec{k}\,]\right\}.
\end{equation}
Here $Z$ is the hadron charge, and its $g$-factor is related as
follows to the magnetic moment $\mu$ measured in the nuclear
magnetons $e/2m_p$:
$$g=\mu/s.$$
In the forward scattering case, when the hadron is at rest
($\vec{p}=0$) and both photons have physical, transverse
polarizations ($(\vec{k}\vec{e})=(\vec{k}\vec{e^{\prime}})=0$), this
vertex reduces to the pure spin interaction. The nonrelativistic pole
scattering amplitude generated by this interaction is
\begin{equation}\label{m1}
M_1=\,M_{1mn}e^{\prime}_m e_n=\left(\frac{e}{2m_p}\right)^2\omega\,g^2\,
i\left( \vec{s}\cdot[\vec{e}\,^{\prime}\times\vec{e}\,] \right) .
\end{equation}
However, this expression is incomplete. Indeed, being applied to a
proton, it does not reproduce the well-known result \cite{lo,gg}
according to which the spin-dependent forward scattering amplitude is
proportional
to $(g-2)^2$. The explanation was pointed out in Ref. \cite{mi}: the
nonrelativistic pole amplitude should be supplemented by a contact
term generated by the spin-orbit interaction, which restores the
agreement with the classical result
\cite{lo,gg}.

This contact term can be easily derived for an arbitrary spin (as
well as the nonrelativistic pole contribution (\ref{m1})) in the
following way. The motion equation for spin in an external electric
field $\vec{E}$ can be written to lowest nonvanishing order in $v/c$
as
\begin{equation}\label{fb}
\frac{d\vec{s}}{dt}\,=\,\frac{e}{2m_p}\,\left(g\,-\,\frac{Z}{A}\right)\,
\left[\vec{s}\times[\vec{E}\times\vec{v}]\right].
\end{equation}
Here $A$ is the target mass as measured in the units of $m_p$ (i.e.,
it is the atomic number in the case of nuclei we are mainly
interested in in the present work.) The interaction Hamiltonian
generating equation (\ref{fb}) is obviously
\begin{equation}\label{h}
H_s \,=\,-\,\frac{e}{2m_p}\left(g\,-\,\frac{Z}{A}\right)\,
\left(\vec{s}\cdot[\vec{E}\times\vec{v}]\right).
\end{equation}
Expressions (\ref{fb}), (\ref{h}) are in fact only slightly rewritten
formulae from book \cite{bl}.  After substituting
$$\vec{v}\,=\,\frac{\vec{p}\,-\,Ze\vec{A}}{2Am_p}$$
into Hamiltonian (\ref{h}), we arrive at the following contact
interaction:
\begin{equation}\label{c}
V_c
\,=\,\left(\frac{e}{2m_p}\right)^2\frac{2Z}{A}\,\left(g\,-\,\frac{Z}{A}\right)\,
\left( \vec{s}\cdot[\vec{E}\times\vec{A}] \right).
\end{equation}
It produces one more piece in the scattering amlitude:
\begin{equation}\label{m2}
M_2=\,M_{2mn}e^{\prime}_m e_n=\,\left(\frac{e}{2m_p}\right)^2\omega
\,\left(-4\,\frac{Z}{A}\right)\,\left(g\,-\frac{Z}{A}\right)\,
i\left( \vec{s}\cdot[\vec{e}^{\prime}\times\vec{e}\,] \right) .
\end{equation}

Taken together, expressions (\ref{m1}) and (\ref{m2}) generate the
forward scattering amplitude:
\begin{equation}\label{m}
M\,=\,\left(\frac{e}{2m_p}\right)^2\omega\,\left(g\,-\,2\frac{Z}{A}\right)^2\,
i\left( \vec{s}\cdot[\vec{e}^{\prime}\times\vec{e}\,] \right) .
\end{equation}
This is the generalization of the low-energy theorem we are looking for.

In the particular case of a proton ($s=1/2,\;Z=A=1$) this formula
reduces to the result of Refs. \cite{lo,gg}.

\section{Low-energy theorem and deuterium hyperfine structure}

Being dependent on nuclear spin, the low-energy amplitude obtained
contributes to the atomic hyperfine structure. However, to apply it
to this problem, the amplitude should be modified. Indeed, both
photons exchanged between the nucleus and electron are
off-mass-shell. So, here $\omega\,\neq\,|\vec{k}|$. Then, virtual
photons have extra polarizations.  We will use the gauge $A_0 =\,0$
where the photon propagator is
\begin{equation}\label{po}
D_{im}(\omega, \vec{k})\,=\,\frac{\,d_{im}}{\omega^2-\,\vec{k}^2},\;\;
d_{im}=\,\delta_{im}-\,\frac{k_i k_m}{\omega^2};\;\;\; D_{00}=\,D_{0m}=\,0.
\end{equation}

Now, first of all, the magnetic moment contribution $M_{1mn}$ to the
pole diagram changes to
\begin{equation}\label{m11}
\tilde{M}_{1mn}\,=\left(\frac{e}{2m_p}\right)^2 g^2\,
i\,\epsilon_{mnk}k_k(\vec{k}\cdot\vec{s})\,\frac{1}{\omega} .
\end{equation}
Second, the convection current, which is proportional to $\pm\vec{k}$
for a nucleus at rest, is operative now and induces the following
nuclear-spin-dependent contribution to the forward scattering
amplitude:
\begin{equation}\label{m3}
M_{3mn}\,=\,-\,\left(\frac{e}{2m}\right)^2 2\,\frac{Z}{A}\,g\,
i\,(k_m\epsilon_{nrs}k_r s_s\,-\,k_n\epsilon_{mrs}k_r s_s)\frac{1}{\omega}.
\end{equation}

We are ready now to write down the electron-nucleus
nuclear-spin-dependent scattering amplitude generated by the two-photon
exchange with the deuteron intermediate state:
\begin{equation}\label{f}
T_{el}=\,4\pi\alpha i\,\int\frac{d^4 k}{(2\pi)^4}\,\frac{d_{im}d_{jn}}{k^4}\,
\frac{\gamma_i(\hat{l}-\hat{k}+m_e)\gamma_j}{k^2-\,2lk}\,(\tilde{M}_{1mn}+M_{2mn}+
M_{3mn}).
\end{equation}
Here $l_{\mu}=\,(m_e,0,0,0)$ is the electron momentum. The structure
$\gamma_i(\hat{l}-\hat{k}+m_e)\gamma_j$ reduces to
$-\,i\,\omega\,\epsilon_{ijl}\sigma_l$ where $\vec{\sigma}$ is the
electron spin.  We will calculate this Feynman integral with
logarithmic accuracy. Two subtle points should be mentioned here. The
singularity at $\omega\,=\,0$, originating from $1/\omega^2$ in the
projection operator (\ref{po}), should be understood in the sense of
the principal value. Then, one comes across a term containing
integral
$$\int_0 \frac{d|\vec{k}|}{|\vec{k}|^2}$$
which diverges in a power-like way at $|\vec{k}|\rightarrow 0$. To
regularize it one should introduce non-vanishing electron velocity
$v$; in this way one obtains the well-known Coulomb wave-function
correction $\pi\alpha/v$ which should be neglected to our purpose. In
our logarithmic approximation one should neglect as well possible
constant terms originating from this integral.

The final result of calculation can be conveniently presented in the
following form. Let us write down the spin-dependent Born term in the
electron-nucleus scattering amplitude:
\begin{equation}\label{b}
T_{0}=\,-\,\frac{2\pi\alpha}{3m_e m_p}\,g\,(\vec{\sigma}\cdot\vec{s}).
\end{equation}
It is in fact minus Fourier-transform of the lowest order contact
hyperfine interaction.  Therefore, the ratio
$\Delta_{el}\,=\,T_{el}/T_0$ is nothing else but the relative
magnitude of the discussed correction to the hf structure. The result
for integral (\ref{f}) can be written as
\begin{equation}
\Delta_{el}=\,\frac{3\alpha}{8\pi}\,\frac{m_e}{m_p}\,\log{\frac{\Lambda}{m_e}}\,
\frac{1}{g}\,
\left(\,g^2\,-\,4g\,\frac{Z}{A}\,-\,12\,\frac{Z^2}{A^2}\right).
\end{equation}

At $s=1/2,\,A=Z=1$ it agrees with the corresponding results
\cite{ar,gy} for muonium (where the effective cut-off $\Lambda$ is
provided by the muon mass) and hydrogen (where the integral is cut
off at the typical hadronic scale $m_{\rho}$).

In the case of deuterium ($s=1,\,g=\mu_d=0.857,\,A=2,\,Z=1$), we are
mainly interested in, the integration over the momentum transfer $k$
is cut off at the inverse deuteron size $\kappa\,=\,45.7$ MeV. In
this way we obtain the following result for the relative correction
in deuterium:
\begin{equation}\label{de}
\Delta_{el}=\,\frac{3\alpha}{8\pi}\,\frac{m_e}{m_p}\,\log{\frac{\kappa}{m_e}}
\left(\mu_d -2-\,\frac{3}{\mu_d}\right).
\end{equation}

At larger momentum transfers, $k>\kappa$, the amplitude of the
Compton scattering on a deuteron is just the coherent sum of those
amplitudes on free proton and neutron. This contribution to the hf
structure can be also easily obtained with the above formulae. Since
both nucleons are in the triplet state, one can substitute
$\vec{s}/2$ both for $\vec{s}_p$ and for $\vec{s}_n$. With the
logarithmic integral cut off here at the usual hadronic scale
$m_{\rho}\,=\,770$ MeV, we get in this way
\begin{equation}\label{di}
\Delta_{in}=\,\frac{3\alpha}{4\pi}\,\frac{m_e}{m_p}\,
\log{\frac{m_{\rho}}{\kappa}}\,\frac{1}{\mu_d}\,
(\,\mu_p^2-\,2\,\mu_p\,-\,3\,+\,\mu_n^2\,).
\end{equation}
Here $\mu_p\,=\,2.79$ and $\mu_n\,=\,-\,1.91$ are the proton and
neutron magnetic moments.

In the conclusion of this section let us mention strong numerical
cancellation between $\Delta_{el}$ and $\Delta_{in}$.

\section{Contribution of deuteron virtual excitations}

The low-energy Compton amplitude discussed above is just linear term
of the full amplitude expansion in $\omega$ (and in
$|\vec{k}|^2/\omega$ for virtual photons). One may expect, however,
that for the deuteron with its small binding energy this
approximation is insufficient even in the atomic problem considered
here. And indeed, we will see below that the deuteron virtual
excitations are far from being inessential to our problem, they
strongly dominate the effect discussed. Since the contribution of
large momentum transfers $k>\kappa$ has been calculated already (see
formula (\ref{di}), we confine now to the region $k<\kappa$. All
calculations below are performed in the zero-range approximation
(zra) for the deuteron which allows to obtain all the results in a
closed analytical form.

Let us start with the transitions induced by the spin
interaction only.  The corresponding scattering amplitude is
$$M_{4mn}\,=\,-\,\left(\frac{e}{2m_p}\right)^2
\sum_n\left\{\frac{<0|[\vec{k}\times\vec{S}]_m|n>
<n|[\vec{k}\times\vec{S}^{\dagger}]_n|0>}{\omega\,-\,E_n\,-I}\right.$$
\begin{equation}\label{ms}
\left.\;-\;\frac{<0|[\vec{k}\times\vec{S}^{\dagger}]_n|n>
<n|[\vec{k}\times\vec{S}]_m|0>}{\omega\,+\,E_n\,+I}\right\}.
\end{equation}
Here $I=\kappa^2/m_p$ is the deuteron binding energy, $E_n=p^2/m_p$
is the energy of the intermediate state $|n>$ (all intermediate
states belong to the continuous spectrum), and
$$\vec{S}=\mu_p
\vec{\sigma}_p e^{i\vec{k}\vec{r}/2}+
\mu_n \vec{\sigma}_n e^{-i\vec{k}\vec{r}/2}$$
where $\vec{\sigma}_{p(n)}$ is the proton (neutron) spin operator.

When calculating this contribution, we will retain only terms
logarithmic in the parameter
$\epsilon\,=\,I/\kappa\,=\,\kappa/m_p\,\ll 1$. The log is generated
by the integration over $k$, and to obtain it it is sufficient to put
the exponents in $\vec{S}$ equal to unity. Then the operator $\vec{S}$ can
induce only M1 transitions.  In the zra the deuteron ground state is
pure $^3 S_1$ from which M1 transition is possible also to $S$-states
only. But due to the orthogonality of the radial wave functions of
different triplet states, the intermediate states confine to $^1
S_0$.

Since the total spin operator
$\vec{s}\,=\,(1/2)(\vec{\sigma}_p\,+\,\vec{\sigma}_n)$ does not induce
triplet-singlet transitions, the operator $\vec{S}$ reduces here to
$$\vec{S}\rightarrow
(\mu_p\,-\,\mu_n)\,\frac{1}{2}\,(\vec{\sigma}_p\,-\,
\vec{\sigma}_n).$$
In our problem of hf structure we need the antisymmetric part of
tensor (\ref{ms}) which is linear in the deuteron spin $\vec{s}$. It
looks now as follows:
\begin{equation}\label{ms1}
M_{mn}^1\,=\,-\,\left(\frac{e}{2m_p}\right)^2\,(\mu_p\,-\,\mu_n)^2\,
i\,\epsilon_{mnk}k_k(\vec{k}\cdot\vec{s})\,
\omega\,\int\frac{d\vec{p}}{(2\pi)^3}\,\frac{|<\,^1S_0,\,p\,|^3 S_1>|^2}
{\omega^2\,-\,(p^2+\kappa^2)^2/m_p^2}
\end{equation}
where $<\,^1S_0,\,p\,|^3 S_1>$ is the overlap integral of the ground
state zra wave function
\begin{equation}\label{gs}
\psi_0\,=\,\sqrt{\frac{\kappa}{2\pi}}\,\frac{e^{-\kappa r}}{r}
\end{equation}
with the singlet one of the momentum $p$.

This contribution to the electron-deuteron scattering amplitude
\begin{equation}\label{in1}
T_{in}^1=\,4\pi\alpha i\,\int\frac{d^4 k}{(2\pi)^4}\,\frac{d_{im}d_{jn}}{k^4}\,
\frac{\gamma_i(\hat{l}-\hat{k}+m_e)\gamma_j}{k^2-\,2lk}\,M_{mn}^1
\end{equation}
is easily calculated with the logarithmic accuracy. Indeed, to this
accuracy the energy denominator in formula (\ref{ms1}) can be
simplified to $$\frac{1}{\omega^2\,-\,\kappa^4/m_p^2}.$$ Then the
integration over $\vec{p}$ reduces to the completeness relation. The
resulting relative correction to the deuterium hf structure is
\begin{equation}\label{di1}
\Delta_{in}^1=\,\frac{3\alpha}{8\pi}\,
\frac{m_e}{m_p}\,\log{\frac{m_p}{\kappa}}\,
\frac{(\,\mu_p\,-\,\mu_n\,)^2}{\mu_d}.
\end{equation}

Let us consider at last the inelastic contribution induced by the
combined action of the convection and spin currents. Since the
convection current is spin-independent, all the intermediate states
are triplet ones as well as the ground state. Therefore, here
operator $\vec{S}$ simplifies to
\begin{equation}\label{si}
\vec{S}\rightarrow \vec{s}(\mu_p e^{i\vec{k}\vec{r}/2}+
\mu_n e^{-i\vec{k}\vec{r}/2}).
\end{equation}
According to the common selection rules, neither $^3 S_1$ can be
excited by the convection current from the ground state. But in the
zra all the states with $l\neq 0$ are free ones. It means that here
we can use as the intermediate states just plane waves, eigenstates
of the momentum operator. Then the only matrix element entering the
amplitude is
\begin{equation}\label{me}
<0\,|e^{\pm i\vec{k}\vec{r}/2}|\vec{p}>\,=\,\frac{\sqrt{8\pi\kappa}}
{(\vec{p}\pm\vec{k}/2)^2\,+\,\kappa^2}.
\end{equation}
In this way the amplitude itself simplifies to
$$M_{mn}^2\,=\left(\frac{e}{2m_p}\right)^2\,
2\,\kappa\,\omega\,\int\frac{d\vec{p}}{\pi^2}\,
\left\{\frac{\mu_p}{[(\vec{p}-\vec{k}/2)^2\,+\,\kappa^2]^2}
+\,\frac{\mu_n}{[(\vec{p}-\vec{k}/2)^2\,+\,\kappa^2][(\vec{p}+\vec{k}/2)^2\,
+\,\kappa^2]}\right\}$$
\begin{equation}\label{ms2}
\cdot\frac{(2p\,-\,k/2)_m\,i\,\epsilon_{nrs}k_r s_s\,
-\,(2p\,-\,k/2)_n\,i\,\epsilon_{mrs}k_r s_s}
{\omega^2\,-\,(p^2+\kappa^2)^2/m_p^2}.
\end{equation}

Integrals we come across when calculating the corresponding part of
the electron-deuteron scattering amplitude
\begin{equation}\label{in2}
T_{in}^2=\,4\pi\alpha i\,\int\frac{d^4 k}{(2\pi)^4}\,\frac{d_{im}d_{jn}}{k^4}\,
\frac{\gamma_i(\hat{l}-\hat{k}+m_e)\gamma_j}{k^2-\,2lk}\,M_{mn}^2,
\end{equation}
are rather tedious even when we confine to terms singular in the parameter
$\epsilon=\kappa/m_p\ll 1:\;1/\epsilon$ and $\log{\epsilon}$.
This relative correction to the hf structure is to this approximation
\begin{equation}\label{di2}
\Delta_{in}^2=\,\alpha\,\frac{m_e}{2\kappa}\,\frac{\mu_p\,-\,\mu_n}{\mu_d}\,-
\,\frac{3\alpha}{\pi}\,\frac{m_e}{m_p}\,\log{\frac{m_p}{\kappa}}\,
\frac{\mu_p\,-\,\mu_n}{\mu_d}.
\end{equation}

One of the terms in this correction,
$$-\,\alpha\,\frac{m_e}{2\kappa}\,\frac{\mu_n}{\mu_d},$$
was obtained and discussed in Refs. \cite{bo,low,los,gf}.
However, another term
$$\alpha\,\frac{m_e}{2\kappa}\,\frac{\mu_p}{\mu_d},$$
is larger numerically.

\section{Corrections due to finite distribution of the deuteron charge
 and magnetic moment}

In the case of hydrogen this problem was considered many years ago \cite{ze}.
For deuterium those corrections should be obviously larger. In the zra
the problem has here a closed solution.

Let us start with the second-order amplitude of the electron-deuteron
scattering as induced by the deuteron charge and magnetic moment. The
nucleus will be treated in the static limit. However, its finite
charge and magnetic moment distributions will be taken into account
by introducing the corresponding form-factors, $F_{ch}$ and $F_m$.
This amplitude is
$$V=-\,(4\pi\alpha)^2\frac{\mu_d}{2m_p}\,\int\frac{d\vec{q}}{(2\pi)^3}\,
i\,[\vec{s}\times\vec{q}\,]\,
\frac{F_{ch}(\vec{q}\,^2)\,F_m(\vec{q}\,^2)}{\vec{q}\,^4}$$
\begin{equation}
\cdot\frac{\gamma_0(\hat{l}+\hat{q}+m_e)\,\vec{\gamma}\,
-\,\vec{\gamma}\,(\hat{l}+\hat{q}+m_e)\,\gamma_0}{(l+q)^2-m_e^2}\;.
\end{equation}
Here again $l_{\mu}=\,(m_e,0,0,0)$, and $q_{\mu}=\,(0,\vec{q})$. This
expression can be conveniently transformed to
\begin{equation}
V=\frac{8m_e\alpha}{\pi}\,\int\frac{dq}{q^2}\,F_{ch}(q^2)\,F_m(q^2)\,T_0
\end{equation}
where $T_0$ is the momentum-independent magnetic Born amplitude (\ref{b}).

The effect we are interested in, vanishes of course if unity is
substituted for both form-factors. Therefore, the corresponding
relative correction to the Born amplitude $T_0$ and to the hf
splitting is in fact
\begin{equation}\label{for}
\Delta_f=\frac{8m_e\alpha}{\pi}\,\int\frac{dq}{q^2}\,
\left[F_{ch}(q^2)\,F_m(q^2)
-1\right].
\end{equation}

In the zra both deuteron form-factors have simple form:
\begin{equation}\label{form}
F_{ch}(q^2)\,=\,F_m(q^2)\,=\,<0|e^{i\vec{q}\,\vec{r}/2}|0>\,=\,\frac{4\kappa}{q}\,
\mbox{arctg}\frac{q}{4\kappa}.
\end{equation}
Substituting it into formula (\ref{for}), we get the following
explicit expression for the correction discussed:
\begin{equation}\label{ch}
\Delta_{f}=\,-\,\alpha\,\frac{m_e}{3\,\kappa}\,(1+2\log2).
\end{equation}

Two closely related features of the effect (in no way specific for
deuterium only) are worth emphasizing here. This correction is of
first (but not second) order in the ratio of the nuclear size to the
Bohr radius $m_e\alpha/\kappa$. Then, contrary to possible naive
expectations, the contributions of the charge and magnetic
form-factors are not additive. Both circumstances can be traced back
to the fact that the typical momenta entering integral (\ref{for})
are of the nuclear, but not atomic, scale.

\section{Deuterium hf structure, discussion of results}
Our total result for the nuclear-structure corrections to the
deuterium hf structure, comprising all the contributions, (\ref{de}),
(\ref{di}), (\ref{di1}), (\ref{di2}), (\ref{ch}), is
$$\Delta=\,\alpha\,\frac{m_e}{2\,\kappa}\,
\left\{\frac{\mu_p\,-\,\mu_n}{\mu_d}\,-\,\frac{2}{3}(1+2\log2)\right\}\,+\,
\,\frac{3\alpha}{8\pi}\,\frac{m_e}{m_p}\,\log{\frac{m_p}{\kappa}}\,
\frac{(\,\mu_p\,-\,\mu_n\,)^2}{\mu_d}$$
\begin{equation}
-\,\frac{3\alpha}{\pi}\,\frac{m_e}{m_p}\,\log{\frac{m_p}{\kappa}}\,
\frac{\mu_p\,-\,\mu_n}{\mu_d}\,
+\,\frac{3\alpha}{8\pi}\,\frac{m_e}{m_p}\,
\log{\frac{\kappa}{m_e}}\,\frac{1}{\mu_d}\,
(\,\mu_d^2\,-\,2\mu_d\,-\,3)
\end{equation}
$$+\,\frac{3\alpha}{4\pi}\,\frac{m_e}{m_p}\,
\log{\frac{m_{\rho}}{\kappa}}\,\frac{1}{\mu_d}\,
(\,\mu_p^2-\,2\,\mu_p\,-\,3\,+\,\mu_n^2\,).$$
Numerically this correction to the hf splitting in deuterium constitutes
\begin{equation}
\Delta \nu\,=\,43\, \mbox{kHz}.
\end{equation}

It should be compared with the lacking $45$ kHz (see (\ref{dif})).
Taking into account the approximations made, first of all the crude
nuclear model (zra), then the neglect of nonlogarithmic
contributions, we believe that the agreement is quite satisfactory.
In particular, including the correction due to the effective
interaction radius $r_0$ into the normalization of the deuteron
ground state wave function (see details in the next section) would
certainly enhance some contributions.

Clearly, the nuclear effects discussed are responsible for the bulk
of the difference between the pure QED calculations and the
experimental value of the deuterium hf splitting. The calculation of
this hf correction, including accurate treatment of nuclear effects,
would serve as one more sensitive check of detailed models of
deuteron structure.

\section{Nuclear polarizability and Lamb shift in deuterium}

The nuclear polarizability contribution to the Lamb shift was
considered recently in Refs. \cite{plh,pwh,lr}. Here we present an
analytical calculation of the effect with a closed result. The zra
approximation used by us is applicable when the region of the wave
function localization is much larger than the interaction range.
Essentially the same condition is necessary to use, instead of the
true interaction, the crude approximation of the square-well
potential, as it is done in Refs. \cite{plh,pwh}.

The effect we are interested in now, is due to the photon-deuteron
scattering amplitude induced by the convection current only. We will
see that in this problem, that of the nuclear polarizability
contribution to the Lamb shift, the characteristic values of the
photon 4-momenta are as follows:
\begin{equation}\label{r}
\omega,\,|\vec{k}|\leq I=\frac{\kappa^2}{m_p}\ll \kappa \sim |\vec{p}|.
\end{equation}
Therefore, now we can omit in the Compton amplitude all dependence on
$\vec{k}$.  As well as in the amplitude $M_{mn}^2$, all the
intermediate states here have $l\neq 0$ and can be described
therefore by plane waves. We will use again matrix element
(\ref{me}), but this time at $\vec{k}\,=\,0$. At last, in the present
problem we are interested in the scalar part of scattering amplitude
which reduces to
$$-\,\left(\frac{e}{2m_p}\right)^2\,
\frac{4}{3}\,\delta_{mn}\,\kappa\,\int\frac{d\vec{p}}{\pi^2}\,
\frac{p^2}{(p^2+\kappa^2)^2}\,
\left\{\frac{1}{\omega\,-\,(p^2\,+\,\kappa^2)/m_p}\,
-\,\frac{1}{\omega\,+\,(p^2\,+\,\kappa^2)/m_p}\right\}.$$
We subtract from the expression in braces the term
$$-\,2\,\frac{m_p}{p^2\,+\,\kappa^2}.$$
After integrating over $\vec{p}$ this term being added to the
Thomson scattering (seagull)
amplitude for a proton, $-\,e^2/m_p$, reproduces the correct one for a
deuteron, $-\,e^2/2m_p$. With the identity
$$\frac{1}{\omega\,-\,u}\,-\,\frac{1}{\omega\,+\,u}\,+\,\frac{2}{u}\,=\,
\frac{2\omega^2}{(\omega^2-\,u^2)\,u}$$
we get the following expression for the photon-deuteron scattering
amplitude in question:
$$M_{mn}^3\,=\,-\,\left(\frac{e}{2m_p}\right)^2\,
\frac{8}{3}\,\delta_{mn}\,\kappa\,\omega^2 m_p\,$$
\begin{equation}\label{}
\cdot\int\frac{d\vec{p}}{\pi^2}\,
\frac{p^2}{(p^2+\kappa^2)^3\,
\{\omega^2\,-\,(p^2\,+\,\kappa^2)^2/m_p^2\}}.
\end{equation}

The contribution of this tensor to the electron-deuteron scattering amplitude
\begin{equation}\label{in3}
T_{in}^3=\,4\pi\alpha i\,\int\frac{d^4 k}{(2\pi)^4}
\,\frac{d_{im}d_{jn}}{k^4}\,
\frac{\gamma_i(\hat{l}-\hat{k}+m_e)\gamma_j}{k^2-\,2lk}\,M_{mn}^3
\end{equation}
can be easily transformed to
$$T_{in}^3=\,\frac{32\pi^2\alpha^2\kappa}{3m_p}
\int\frac{d\vec{p}}{\pi^2}\,\frac{p^2}{(p^2+\kappa^2)^3}$$
\begin{equation}\label{t}
\cdot i\,\int\frac{d^4 k}{(2\pi)^4}\,\left\{\frac{1}{\omega\,(k^2-\,2lk)}\,
+\,\frac{2\,\omega^3}{k^4\,(k^2-\,2lk)}\right\}\,
\frac{1}{\omega^2\,-\,(p^2\,+\,\kappa^2)^2/m_p^2}.
\end{equation}
The first term in the braces here contains no photon propagation,
neither $1/k^4$, nor $1/k^2$. In other words, it corresponds to the
instantaneous Coulomb interaction. The second term corresponds to the
exchange by three-dimensionally transverse quanta, i.e., to the
magnetic interaction of convection currents.

Perhaps, the most convenient succession of integrating expression
(\ref{t}) is as follows: the Wick rotation; transforming the integral
over the Euclidean $\omega$ to the interval $(0,\,\infty)$; the
substitution $\vec{k}\rightarrow \vec{k}\,\omega$; integration over
$\omega$; integration over $\vec{k}$ (at the last two procedures it
gets clear that the effective values of $\omega,\,|\vec{k}|$ belong
to interval (\ref{r}) indeed); at last, integration over $p$. The
following identity is useful here:
$$\int_0^1dx\,(1-x)^{a-1} x^{b-1} \log x \,=
\,\frac{\Gamma(a)\,\Gamma(b)}{\Gamma(a+b)}\,[\,\psi(b)\,-\,\psi(a+b)];\;\;
\psi(b)\,=\,\frac{d}{db}\log{\Gamma(b)}.$$

The effective electron-nucleus interaction operator (equal to
$\,-\,T_{in}^3$) can be finally presented in the coordinate
representation as
\begin{equation}\label{vle}
V_{le}\,=\,-\alpha\,m_e\,\alpha_d(0)\,5 \left(\log\frac{8I}{m_e}\,+
\,\frac{1}{20}\right)\delta(\vec{r}).
\end{equation}
Here $\alpha_d(0)$ is the static value of the deuteron electric
polarizability defined as usual by the relation
\begin{equation}\label{ad}
\alpha_d(\omega)\,=\,4\pi\alpha\,\frac{2}{3}\,\int\frac{d\vec{p}}{(2\pi)^3}\,
\frac{p^2+\kappa^2}{m_p}\;
\frac{<0|\vec{r}\,|n><n|\vec{r}\,|0>}
{(p^2\,+\,\kappa^2)^2/m_p^2\,-\,\omega^2}.
\end{equation}
The matrix elements entering expression (\ref{ad}) are dominated by
large distances. In this asymptotic region the naive zra expression
(\ref{gs}) for the deuteron ground state wave function should be
supplied by the correction factor $\;(1\,-\,r_0\kappa)^{-1/2}\;$
taking into account finite effective interaction radius $r_0$ (see
Refs. \cite{bl,ff}). In this way we get the following result for the
static electric polarizability:
\begin{equation}
\alpha_d(0)\,=\,\frac{\alpha}{32\,(1\,-\,r_0\kappa)}\,\frac{m_p}{\kappa^4}\,
=\,0.64\,\mbox{fm}^3.
\end{equation}
This numerical value is close to the experimental one \cite{rk}:
$0.70(5)\,\mbox{fm}^3$ (as well as to the values
$0.613,\,0.623,\,0.625\,\mbox{fm}^3$ obtained in Ref. \cite{lr} with
different separable nuclear potentials and to $0.635\,\mbox{fm}^3$
found in Ref. \cite{plh} with a square-well potential).

The overall result (\ref{vle}) consists of two contributions of
different physical origin. The dominating one is generated by the
instantaneous Coulomb interaction. Its contribution to the overall
numerical factor
$$-5\left(\log\frac{8I}{m_e}\,+\,\frac{1}{20}\right)$$
in formula (\ref{vle}) is
$$-4\left(\log\frac{8I}{m_e}\,+\,\frac{5}{12}\right).$$
The magnetic interaction contributes to the overall factor
$$-\left(\log\frac{8I}{m_e}\,-\,\frac{17}{12}\right).$$

The level shift of the deuterium ground state produced by operator (\ref{vle})
constitutes\newline
$-\,22.3$ kHz. The Coulomb and magnetic contributions to it are,
respectively, $-\,19.7$ and\newline
$-\,2.6$ kHz. The results are close to
those of Refs. \cite{plh,pwh,lr}.

No wonder that the Coulomb contribution is negative: this is a true
second-order (in the electron-nucleus static interaction) correction
to the ground state of a system consisting of an electron at rest and
a nucleus which is in the ground state itself. The sign of the
magnetic contribution cannot be fixed in this way: in the language of
the common noncovariant perturbation theory this is a fourth-order
correction, second-order in the photon-electron interaction and
second-order in the photon-nucleus one.

One more contribution to the Lamb shift in deuterium is caused by the
deuteron magnetic polarizability, considered earlier also in Ref.
\cite{lr}. This is in fact the contribution of the scalar part of
amplitude (\ref{ms}). The calculation simplifies due to the
following circumstances. First, the numerators $d_{im}$ of the photon
propagators reduce here obviously to $\delta_{im}$. Then, the
integration over $\vec{k}$ is spherically-symmetric one. So, to our
purpose the scalar part of amplitude (\ref{ms}) may be simplified to
$$M_{mn}^4\,=\,-\,4\pi\alpha\,\frac{(\mu_p\,-\,\mu_n)^2\,\kappa\,
(\kappa\,+\,\kappa_1)^2}{9\,m_p^3}$$
\begin{equation}
\cdot\delta_{mn}\,\vec{k}^2\,
\int\frac{d\vec{p}}{\pi^2}\,
\frac{1}{(p^2+\kappa^2)\,(p^2+\kappa_1^2)
\,[\omega^2\,-\,(p^2\,+\,\kappa^2)^2/m_p^2]}.
\end{equation}

We have used here the explicit form of the $^1 S_0$ coordinate wave
function in the zra:
\begin{equation}
\psi_s\,=\,\frac{\sin(pr+\delta)}{\sqrt{2}\,\pi r}
\end{equation}
where
$$\mbox{ctg}\,\delta\,=\,\frac{\kappa_1}{p},\;\;\kappa_1\,=\,7.9 MeV.$$
Its overlap with the ground state zra wave function (\ref{gs})
constitutes
\begin{equation}
<\,^1S_0\,|^3 S_1>\,=\,\frac{\sqrt{8\pi\kappa}\,(\kappa\,+\,\kappa_1)}
{(p^2\,+\,\kappa^2)\sqrt{p^2\,+\,\kappa^2_1}}.
\end{equation}

Further calculations are close to those related to the electric
polarizability; only the last integration, that over $p$, is done
numerically for the nonlogarithmic contribution. The resulting
effective electron-nucleus interaction operator can be written as
\begin{equation}\label{vlm}
V_{lm}\,=\,\alpha\,m_e\,\beta_d(0)\left(\log\frac{8I}{m_e}\,-
\,1.24\right)\delta(\vec{r}).
\end{equation}
Here $\beta_d(0)$ is the static value of the deuteron magnetic polarizability
equal to
\begin{equation}
\beta_d(0)\,=\,\frac{\alpha\,(\mu_p\,-\,\mu_n)^2}
{8\,m_p\,\kappa^2}\,
\frac{1+\kappa_1/3\kappa}{1+\kappa_1/\kappa}.
\end{equation}
This contribution to the Lamb shift of the deuterium ground state
constitutes $0.31$ kHz which is very close to the result of
Ref. \cite{lr}.

\bigskip
\bigskip
We are grateful to M.I. Eides, H. Grotch and M.I. Strikman
for useful discussions. This investigation was
financially supported by the Program "Universities of Russia", Grant
No. 94-6.7-2053.

\end{document}